# Untargeted Effects in Organic Exciton-Polariton Transient Spectroscopy: A Cautionary Tale

Scott Renken[1,†], Raj Pandya[2,3,†], Kyriacos Georgiou[4,5], Rahul Jayaprakash[4], Lizhi Gai[6,7], Zhen Shen[7], David G. Lidzey[4], Akshay Rao[2], Andrew J Musser[1]*

[1] Department of Chemistry and Chemical Biology, Cornell University, Ithaca, NY, USA

[2] Cavendish Laboratory, University of Cambridge, J.J. Thomson Avenue, CB3 0HE, Cambridge, UK

[3] Laboratoire Kastler Brossel, École Normale Superiéure-Université PSL, CNRS, Sorbonne Université, College de France, Paris 75005, France

[4] Department of Physics and Astronomy, University of Sheffield, Sheffield, UK

[5] Department of Physics, University of Cyprus, P.O. Box 20537, Nicosia, 1678, Cyprus

[6] Key Laboratory of Organosilicon Chemistry and Material Technology, Ministry of Education, Hangzhou Normal University, Hangzhou, 311121, China

[7] State Key Laboratory of Coordination and Chemistry, School of Chemistry and Chemical Engineering, Nanjing University, Nanjing, 210046, China

[†] Contributed equally

* ajm557@cornell.edu

**Abstract**

Strong light-matter coupling to form exciton- and vibropolaritons is increasingly touted as a powerful tool to alter the fundamental properties of organic materials. It is proposed that these states and their facile tunability can be used to rewrite molecular potential energy landscapes and redirect photophysical pathways, with applications from catalysis to electronic devices. Crucial to their photophysical properties is the exchange of energy between coherent, bright polaritons and incoherent dark states. One of the most potent tools to explore this interplay is transient absorption/reflectance spectroscopy. Previous studies have revealed unexpectedly long lifetimes of the coherent polariton states, for which there is no theoretical explanation. Applying these transient methods to a series of strong-coupled organic microcavities, we recover similar long-lived spectral effects. Based on transfer-matrix modelling of the transient experiment, we find that virtually the entire photoresponse results from photoexcitation effects other than the generation of polariton states. Our results suggest that the complex optical properties of polaritonic systems make them especially prone to misleading optical signatures, and that more challenging high-time-resolution measurements on high-quality microcavities are necessary to uniquely distinguish the coherent polariton dynamics.

**Introduction**

There has long been substantial interest in using light-matter interactions to alter the photophysical dynamics of molecular materials. Historically, the principal approach has used intense light sources to stimulate reaction intermediates or coherently control the evolution of photoexcited wavepackets through carefully tailored optical pulses.[1–9] Such experiments typically rely on high light intensities and operate in a regime where the material excitations and photons are weakly coupled. Recently, polaritons have emerged as alternative approach to achieve such control in the absence of strong exciting field.[10,11,20,21,12–19] These are hybrid states with mixed photonic and material character, and they are formed from the strong coupling between confined optical fields – often within Fabry-Perot microcavities or in the near-field of a plasmonic surface – and molecular vibrational or electronic absorption transitions. The mixed character of their wavefunctions and straightforward energetic tunability offer the potential to rewrite potential energy landscapes and redirect photophysical pathways.[19,20,22–26] Already, vibrational polaritons have yielded surprising changes in bulk chemical reactivity and rapid intermolecular vibrational energy transfer.[27–30] Similarly, exciton-polaritons formed from strong coupling to electronic absorption transitions have been reported to result in significantly enhanced charge-carrier transport,[31] long-range energy transfer,[32–34] and changes to fundamental singlet-triplet interconversion dynamics from the picosecond to microsecond timescales.[35–38]

In the steady state, polaritons are chiefly identified through angle-dependent reflectivity. Polaritons inherit some of the dispersion from their parent photonic state, yielding angle-dependent upper and lower polariton bands which anti-cross at the parent exciton energy. This anticrossing is the hallmark of the strong coupling regime, and its magnitude defines the Rabi splitting which is used to benchmark the light-matter interaction strength.[39] The dispersion of the lower polariton can additionally be observed in photoluminescence spectroscopy.[40] Following non-resonant excitation (i.e. at energies above the lower polariton), the distribution of emission intensity along the lower polariton dispersion, and thus the energetic distribution of polariton population, depends sensitively on the energetic separation of the cavity photon and exciton. This dependence is widely interpreted to reflect the relaxation dynamics from uncoupled intracavity states (e.g., relaxed excitons, excimers) into the lower polariton and bottlenecks where this relaxation becomes ineffective.[41–46] Such measurements reveal that the principal states within these structures are not polaritons but rather uncoupled 'dark' states. The same can be deduced from first principles; within the Tavis-Cummings model, for N dipoles that couple to the photonic mode, there will result 2 bright states—the polaritons—and N-1 dark states. In bulk organic microcavities, the resulting dark states are expected to predominate by a factor $\sim 10^5$.[47–49] However, their role in polariton-induced modification of photophysics is unclear, not least because they cannot be directly measured by photoluminescence spectroscopy.

Transient absorption spectroscopy provides a useful window into these dynamics, as it is sensitive to bright and dark electronic states. In the first such investigation, Virgili *et al.* probed the dynamics in high-Q cavities containing a J-aggregated dye following excitation with 15 fs pulses.[50] The resulting spectra were interpreted as signatures of excited upper and lower polaritons and uncoupled intracavity excitons, with the kinetics suggesting the dark states mediate relaxation between polariton manifolds. The ability to resolve unique signatures of these states and track their evolution makes this method the most direct route to understand whether and how photophysical pathways are altered by strong coupling. Accordingly, several subsequent studies using low-Q metallic cavities have reported intriguingly long polariton lifetimes (>1 ps) and alterations to intrinsic photochemical processes from energy transport to singlet fission.[33,36,51–56] Crucially, the understanding of the dynamics in these experiments depends sensitively on the ability to robustly assign the observed spectral features to specific electronic states. In the studies on low-Q cavities, almost the entirety of the photoexcited response was assigned to photoexcited polaritons, with no direct signatures of the dark states reported. Parallel studies of vibrational polaritons have similarly found that the most prominent spectral features are centered at the polariton resonances.[28,57–60] However, in vibrational systems it has been shown that these responses can arise from optical artefacts and non-specific effects of photoexcitation.[57–60] The leading example is 'polariton contraction' in which photoexcitation removes some molecules from the ground state and thus reduces the total absorption responsible for polariton formation.[58,59] This should result in a slight dynamic reduction of the Rabi splitting. While clearly identified in vibro-polariton studies and implicated in the blueshifts of organic polariton condensates,[61] it has been considered and explicitly discarded in the transient spectroscopy of exciton-polaritons.[52,62] Recent work by Liu et al. on exciton-polaritons suggests that other effects—the modulation of molecular excited-state absorption spectra by the cavity structure,[62] and carrier heating/thermalization dynamics within metal mirrors[63]—should likewise be taken into account.

Here, we build on these ideas and explore how non-targeted side effects of photoexcitation can alter the transient absorption spectra of exciton-polariton systems, independently of material choice. We perform transient measurements on a set of low- and high-Q microcavities in the strong-coupling regime and observe long-lived, highly structured spectral responses analogous to earlier reports of photoexcited polaritons.[36,50–52] We find that these signatures persist regardless of the choice of excitation wavelength, even in conditions when no polaritons or molecular electronic states should be excited. Instead, we can describe the critical characteristics of these features using a simple model that only considers polariton contraction, thermal expansion, and bulk refractive index changes. Moving beyond structures based on metallic mirrors,[62,63] we find that these optical effects do not require direct photoinduced carrier heating within the mirror and can be similarly prominent in dielectric structures. Moreover, the richer spectrum of dielectric microcavities provides numerous additional signatures of non-specific photoinduced effects. Our model suggests that these features inevitably appear near any optical resonances in the ground state due to the complex optical structure of strong-coupled cavities. In contrast to previous studies,[52,62] we find that within practical pump intensity ranges the shape of these features remains constant, and hence power dependence may not be applied to rule them out, but they may be disentangled through their unique angular responses. Only when measuring with temporal resolution shorter than the very short expected polariton lifetime do we detect signatures that cannot be accounted for in the phenomenological model. Considering that, in the exciton-polaritons field, this stringent condition has only ever been met in the first transient absorption study,[50] these results call for careful re-evaluation of the dynamics of organic exciton-polaritons.

**Methods**

The organic dye BODIPY-R was synthesized following published protocols.[64] For thin film and microcavity preparation, we dissolved the dye to a concentration of 2.5 mg/mL in toluene. To increase processability and minimize aggregation and quenching effects, the dye was co-dissolved with polystyrene (PS, molecular weight 192000, Sigma Aldrich) at a concentration of 25 mg/mL, corresponding to ~10% dye in PS by weight. This solution was spin-coated onto quartz-coated glass slides, thin coverslips or the bottom mirrors of Fabry-Perot microcavities. Two types of microcavity were prepared, using metallic (Ag) or dielectric ($SiO_2$, $TiO_2$) mirrors. Ag was deposited using thermal evaporation, with typical bottom and top mirror thicknesses of 200 nm and 25 nm. $SiO_2$ and $TiO_2$ were deposited with e-beam evaporation, with layer thicknesses of 105 nm ($SiO_2$) and 71 nm ($TiO_2$) in the structure $TiO_2/(SiO_2/TiO_2)_5$ to yield distributed Bragg reflectors (DBRs) with a stop-band centred at 650 nm. Top and bottom DBRs were fabricated using identical evaporation parameters. In both structures, top mirrors were deposited directly onto the spin-coated organic layer, the thickness of which was tuned in the range ~215-240 nm for DBR structures and ~125-140 nm for Ag-Ag structures to yield λ and λ/2 cavities, respectively.

The angle-dependent reflectivity of DBRs and microcavities was characterized on a broadband fiber-coupled goniometer. The resulting reflectivity dispersions were simulated with transfer matrix methods,[65] using as inputs the normal-incidence transmission spectra of a 156-nm BODIPY-R/PS thin film (Figure 1a) and a 7-pair DBR. Following benchmarking using the steady-state reflectivity, a transfer matrix model was used to simulate pump-induced changes to the microcavity optical properties. Time-resolved measurements were performed on a home-built broadband transient absorption spectrometer in both transmission and reflectance geometry <5 °. Two modes of excitation were applied, namely (i) Narrow-band (10 nm FWHM) excitation pulses throughout the visible spectrum were generated with a commercial OPA, affording a temporal resolution of ~200 fs, and (ii) Compressed broadband pulses spanning 520-600 nm or 760-900 nm were used to achieve temporal resolution <15 fs in a similar manner to that detailed by Liebel *et al.*[66] Likewise, two detection systems were employed. (i) For measurements out to long time-delays, we utilized a probe detection range of 500-800 nm. (ii) Variable pump-wavelength measurements were performed on a different system compatible with <15 fs pulses and enhanced probe sensitivity 800-1300 nm but reduced sensitivity <580 nm. The white-light probe continuum spanned 540-1300 nm, though the low transmission through optical microcavities strongly attenuated this range in the visible spectral region.

**Results**

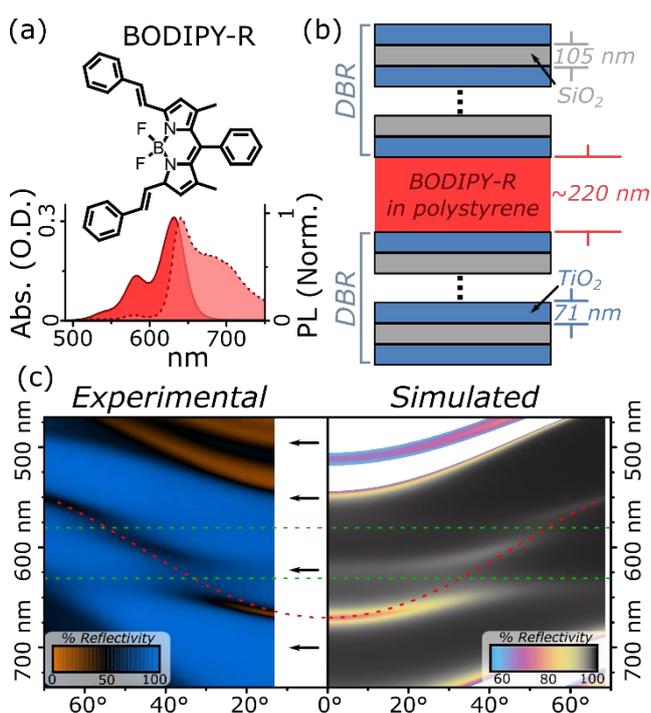

*Figure 1. BODIPY-R microcavities.* **(a)** Chemical structure of BODIPY-R and steady-state optical spectra of 156 nm film of BODIPY-R loaded at 10 wt% in polystyrene matrix. **(b)** Structure of DBR microcavity, consisting of 5.5 pairs of $TiO_2/SiO_2$ alternating layers. $TiO_2$ layers are 71 nm thick and $SiO_2$ layers are 105 nm thick, yielding a stopband centred at 650 nm. Alternative structures used Ag mirrors (top: 25 nm, bottom: 200 nm) instead of DBRs. Organic layer thickness was varied to tune the position of the cavity mode. **(c)** Angle-dependent reflectivity of DBR cavity, revealing two branches defined as lower and upper polaritons, with an anticrossing at the exciton peak at 630 nm. An additional interaction at 582 nm causes slight modulation of the upper polariton dispersion but no additional anticrossing. Transfer matrix model (right) reproduces the full cavity structure. Dashed lines are unperturbed cavity mode and exciton energies. Arrows denote excitation wavelengths applied in transient absorption.

The absorption and photoluminescence of the BODIPY-R/PS thin film show no sign of significant aggregation or excimer emission,[67] and the prominent 0-0 and 0-1 absorption peaks are well resolved at 10% loading. As in numerous other BODIPY derivatives and previous studies of BODIPY-R, at this concentration the dye exhibits sufficient oscillator strength to undergo strong light-matter coupling within metallic and dielectric cavities (Figure 1b).[42,61,68–72] Figure 1c shows typical angle-resolved reflectivity from such a DBR cavity. In addition to the strong dispersion of the stop band, we resolve distinct local minima in the reflectivity surrounding the bare 0-0 vibronic peak at 630 nm. These anti-cross as the angle increases, with a separation of 111 meV demonstrating the achievement of strong coupling.[73] By contrast, at the peak of the 0-1 vibronic absorption at 582 nm, we observe a clear broadening of the reflectivity dispersion and a faint splitting best resolved in the transfer matrix simulation, but no distinct anti-crossing. We thus consider the 0-1 peak to be in the weak or intermediate coupling regime. Accordingly, we label the states detected at lower angles as the lower polariton (LP) and upper polariton (UP). We note that the upper vibronic band, despite not being in the strong-coupling regime, nonetheless shows appreciable interactions which can complicate the analysis of transient data, see below. The full angle-dependent reflectivity can be described using our transfer matrix simulations, confirming our assignments of strong and intermediate light-matter coupling. Applying the model to 'empty' cavities containing only the PS matrix, we obtain cavity Q-factors in our structures of 334 (DBR) and 15 (Ag), corresponding to cavity photon lifetimes of 110 fs and 5 fs which should severely limit the observable polariton dynamics.

To understand the dynamics of the UP and LP following photoexcitation, we apply standard pump-probe spectroscopy techniques, first using state-selective narrow-band excitation. The transient transmission data of the reference BODIPY-R/PS thin film in Figure 2a shows prominent ΔT/T>0 bands at 580 nm, 640 nm and 710 nm, indicating bleaching of the ground-state absorption (GSB) and stimulated emission (SE) from the photoexcited singlet state. Photoinduced absorption (PIA) is characteristically weak in BODIPY dyes and not evident on this scale.[67,74] No changes in spectral shape are evident over the measured range, indicating that no further electronic states are required to describe the intrinsic BODIPY-R/PS film photophysics.

In the microcavities, a seemingly much more complex picture emerges despite the relatively simple electronic structure observed in steady-state reflectivity. As previously highlighted in work on J-aggregate microcavities, it is in principle necessary to measure both transient transmission and reflectance spectra in order to describe the full evolution of the states, since e.g. a decrease in transmission could be correlated with either an increase in excited-state absorption or a transient shift in the ground-state reflectivity.[51] We have thus measured the transient transmission and reflectance of a dielectric cavity containing BODIPY-R, henceforth denoted BODIPY-R/DBR, under identical excitation conditions, shown in Figures 2b,c. In both detection modes we observe qualitatively similar features: sharp derivative-like combinations of positive and negative bands, with each pair centred around the polariton resonances evident in the ground state (650 nm and 610 nm at normal incidence) and opposite sign between transmission and reflectance. Similar features have previously been assigned as unique signatures of excited polariton states.[50–52,75,76] Notably, these signatures persist into the 10's and even 100's of ps timescale, despite the polariton lifetime being nominally limited to <200 fs by the short photonic lifetimes intrinsic to our cavities.[47,77] This paradoxical behaviour has been reported in multiple systems but never explained.[51–53,75,76] We note that the transient transmission and reflectance data contain spectral signatures arising from the same parent states (based on spectral position) and exhibit a similar combination of dynamics spanning the sub-ps to 100-ps scale (Figure 2d). We thus consider them to report on the same physics and henceforth consider only the transmission data.

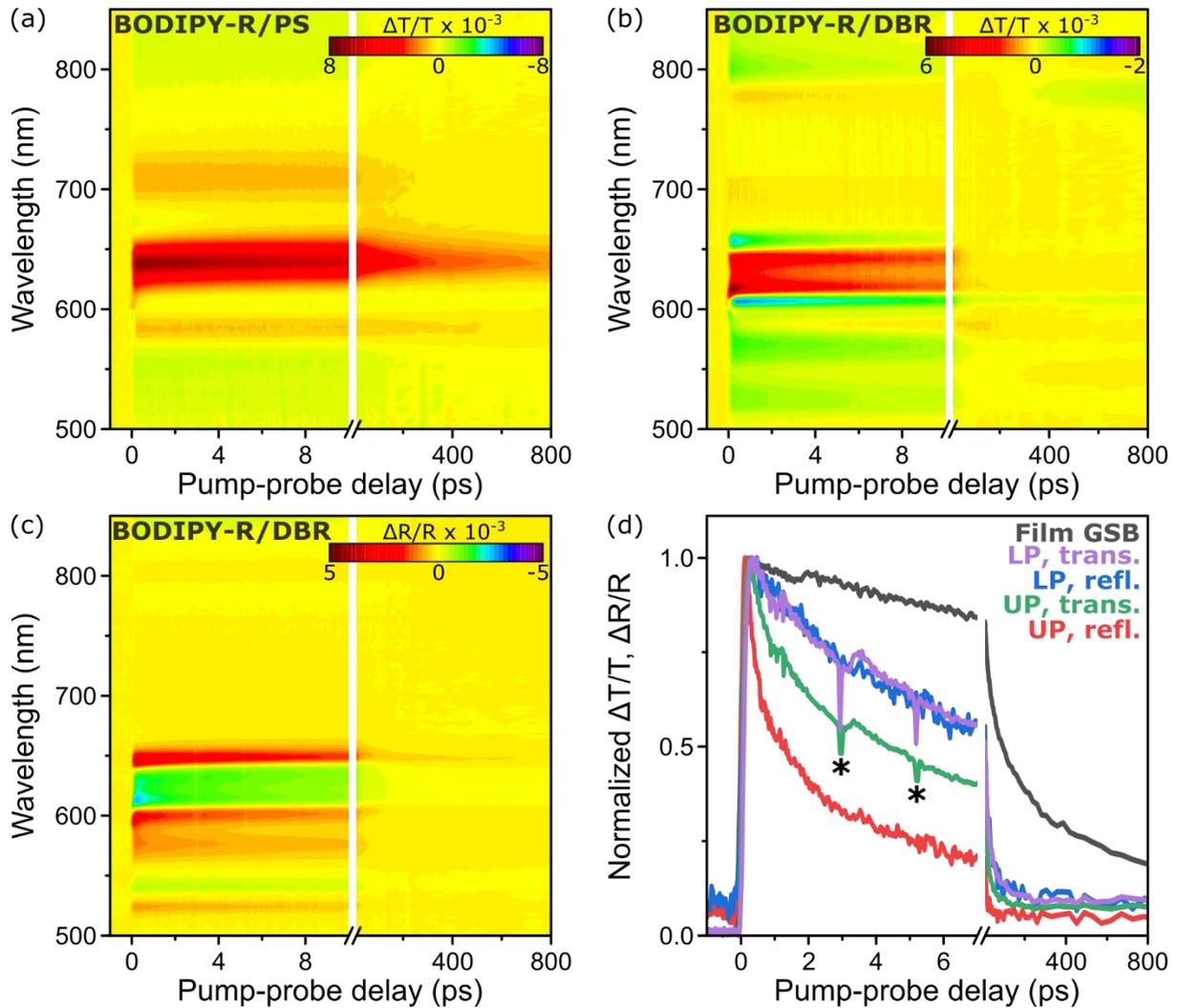

**Figure 2. Transient spectroscopy of BODIPY-R and its dielectric cavities**. (a) Transient transmission of BODIPY-R/polystyrene film, following excitation at 490 nm. **(b)** Transient transmission of DBR cavity, following excitation at 490 nm. **(c)** Transient reflectance of the same cavity in the same conditions. **(d)** Decay kinetics integrated over spectral features indicated in (a)-(c). GSB: ground-state bleach, LP: lower polariton, UP: upper polariton. Sharp dips denoted by asterisks are from pump intensity fluctuations.

To help disentangle the origin of the transient transmission features in Figure 2b, we performed the same experiment with narrowband excitation at 490 nm, 550 nm, 620 nm and 700 nm, and the corresponding spectra at selected time delays are presented in Figure 3a-d. The first two wavelengths are not resonant with any polariton states and, because they fall outside the stop-band of the cavity, are readily absorbed by the BODIPY-R film, generating intracavity excitons. Excitation at 620 nm is approximately resonant with the UP state, though photons transmitted through the top mirror would also be strongly absorbed by the 0-1 absorption transition of BODIPY-R. Excitation at 700 nm is sub-resonant, falling below the dye absorption tail and even the LP, though we cannot rule out excitation into weakly absorbing states within the tail.[69] Crucially, the fundamental behaviour is the same across these conditions: transient spectra with well-defined positive and negative peaks that appear at the same positions, do not shift with time delay and exhibit significant magnitude out to beyond 10 ps, a timescale difficult to reconcile with standard models of exciton-polaritons.[47,77] The features that identifiably recur between experiments are indicated with arrows for clarity.

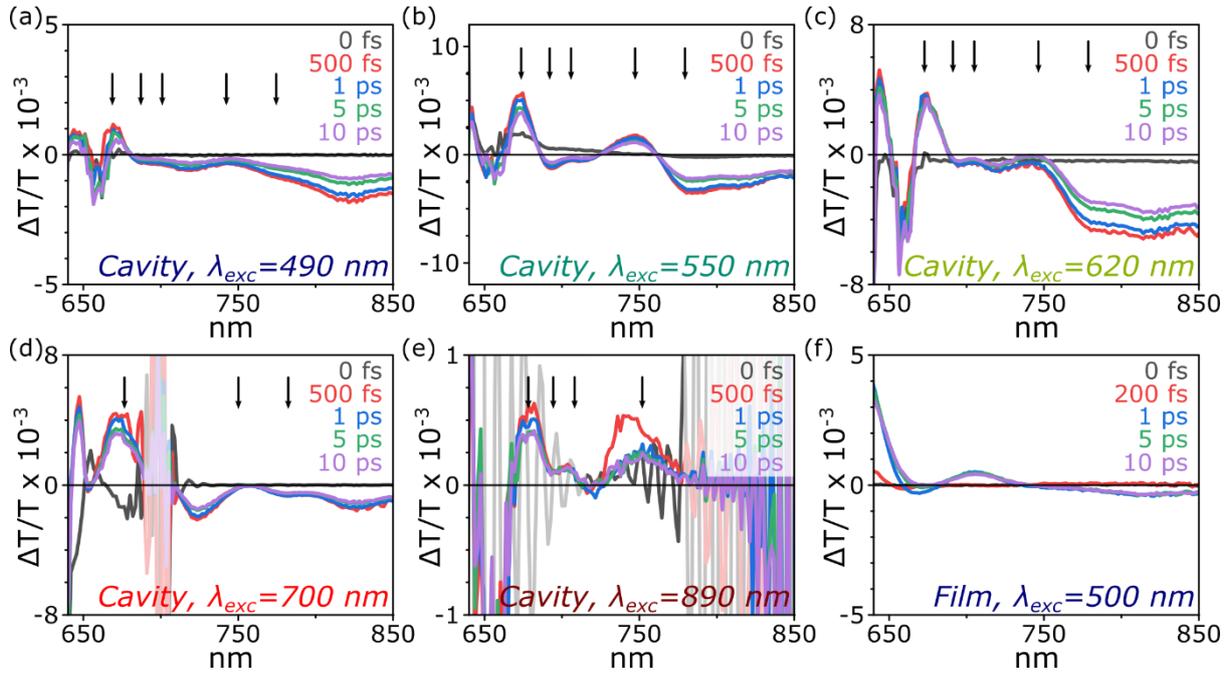

**Figure 3. Pump wavelength dependence of transient transmission.** Transient transmission spectra of DBR cavity following narrowband excitation at **(a)** 490 nm, **(b)** 550 nm, **(c)** 620 nm and **(d)** 700 nm, and **(e)** broadband excitation centred at 890 nm. Significant pump scatter and interference in the latter panels are shaded for clarity. **(f)** Equivalent data from BODIPY-R/PS thin film, excited at 500 nm. Arrows indicate positions of spectral features reproduced between excitation conditions.

Of these features, the sharply structured bands <660 nm coincide with the spectral position of the UP and LP and are consistent with previous assignments to photoexcited UP and LP states.[50–52,75] We note, however, that these are highly correlated with additional features at longer wavelengths, such as positive bands at 675 nm and 750 nm and a negative peak at 770 nm. These cannot be easily linked to the polariton structure but show reasonable spectral agreement with the BODIPY-R stimulated emission (675 nm) and the edge of the stopband and first Bragg mode of the DBR (~750 nm and 770 nm at normal incidence). The persistence of these features, with the same dynamics as in the UP/LP region, raises concern that the transient transmission data includes contributions from effects besides the electronic population of interest. Indeed, the excitation at 700 nm should not be able to generate intracavity excitons or LP states; we ensure that pump intensities are low enough to rule out multi-photon excitation processes. Nonetheless, the same key spectral features can be detected outside of the pump scatter window. To rule out any contributions from excitation of weakly absorbing tail states, we repeated the experiment with a broadband sub-gap pump pulse centred at 890 nm (Figure 2e). Here again we observe the same underlying spectral structure in the middle of our detection window, namely the peaks at 665 nm and 750 nm with finer modulations 685-730 nm, in the absence of any electronic excitations. Strong pump scatter and low sample transmission prevent detection of other signatures.

The principal difference between excitation conditions is the relative magnitude of the underlying photoinduced absorption. This band extends well into the near-infrared, is strongest for excitation most resonant with the BODIPY-R absorption and can be attributed to intra-cavity singlet states from its agreement with the BODIPY-R PIA (Figure 3f). This band provides a useful indication of whether BODIPY-R excitons have been generated, and its absence confirms that no electronic excitations are present in the 890 nm experiment. It is noteworthy that on the observed timescales the BODIPY-R film exhibits negligible spectral evolution whereas <10 ps decay is easily evident in microcavities when electronically excited states can be generated, demonstrating the direct role of the cavity structure in the detected dynamics.

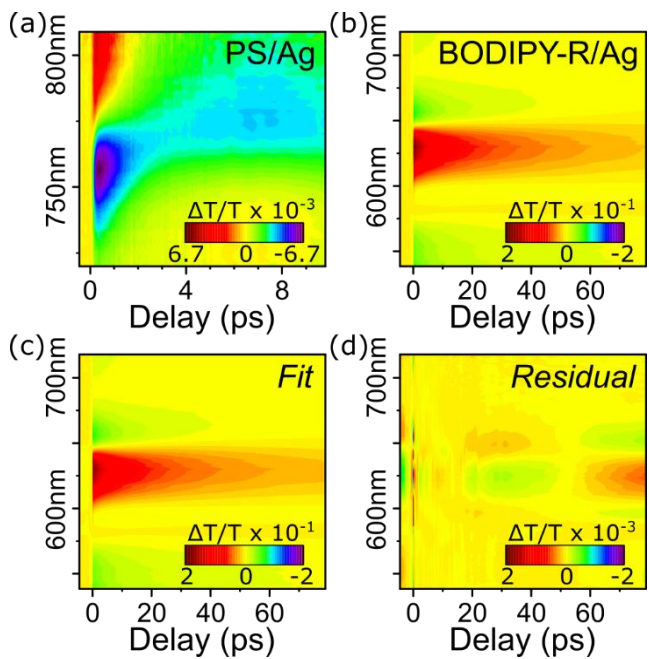

**Figure 4. Cavity transient transmission. (a)** Transient transmission of a metallic 'empty' cavity containing only polystyrene, excited at 550 nm. Long-lived transient response is observed centred on the cavity mode despite the absence of photoexcited states. No dependence on pump wavelength is observed. **(b)** Transient transmission of metallic BODIPY-R cavity taken under similar conditions. **(c)** Global multiexponential fitting can accurately describe the principal decay components. **(d)** The residual following subtraction of the exponential fit reveals slow oscillations around the bare cavity mode. Similar results are obtained with DBR cavities.

We further probed the possible contribution to our spectra of effects other than electronic excitations using a metallic optical cavity containing PS host matrix but no dye molecules. As shown in Figure 4a, upon photoexcitation we observe a sharp derivative-like feature centred on the cavity-mode resonance, which decays on the few-ps timescale to yield a longer-lived photoinduced absorption, despite there being no molecules to photoexcite. These data highlight that no polaritons or electronic excitations are required to produce the types of spectral signatures often attributed to polaritons or apparent dynamics in the few- to 10-ps timescale. Indeed, the transient response of such purely photonic structures is well characterized.[78,79] Here, the dominant effect is likely pump-induced changes to the refractive index of the metallic mirrors, similar to suggestions using TIPS-pentacene microcavities.[63] Similar effects can be obtained within organic films, as reported for instance from a PMMA waveguide mode not coupled to any electronic excitations.[80] We further highlight the pronounced growth kinetics observed in the PIA at 775 nm. These may potentially be related to the coherent phonon oscillations within the Ag mirror,[63] but we additionally note that photoexcitation has been reported to create localized strain in thin films, resulting in propagating strain waves that manifest as time-domain oscillations in the transmission signal.[81–84] These effects can be difficult to detect when superimposed with bulk electronic signatures. However, in a metallic BODIPY-R microcavity we can well describe the principal decay dynamics of the prominent transient bands with a global multi-exponential fit, and the residual shows evidence for a weak, slowly oscillating component (Figure 4b-d). These results suggest that to fully understand the transient transmission spectra of BODIPY-R microcavities in the strong-coupling regime, we must explicitly consider the role of non-polaritonic effects. We draw analogy here to the spectroscopy of organic thin films, where pump-induced, local thermal modification of the ground-state absorption can result in surprisingly long-lived spectral signatures that must be carefully distinguished from the electronic species of interest, especially in the region of sharp ground-state absorption bands.[85–87]

Even in the absence of electronic excitation, non-resonant interactions with the exciting pulse are sufficient to generate coherent vibrations, i.e. deposit energy in the electronic ground state.

To understand the possible contributions of these effects, we turn to our transfer matrix model of the microcavity optical properties. Our transfer matrix model generates the full reflectance, transmittance and absorbance of the microcavity over the full spectral and angular range, based on the optical properties of the constituent layers (see Figure 1b). We seek to replicate the conditions of our transient optical experiments, in which the signal is based on the difference between 'pump off' and 'pump on' measurements. Accordingly, we describe these two conditions by varying key optical parameters within our transfer matrix description of strong-coupled BODIPY-R/DBR and BODIPY-R/Ag cavities. The 'pump off' microcavity is derived from our description of the steady-state reflectivity data in Figure 1. For the pump-modified microcavity, we alter the optical properties through a range of non-specific pump-induced effects, with no explicit treatment of photoexcited electronic states. To convert these results into a proxy transient transmission signal, we calculate $\frac{\Delta T}{T} = \frac{T_{sim\,ON} - T_{sim\,OFF}}{T_{sim\,OFF}}$ at each wavelength and angle. For example, to capture the potential effects of thermal expansion, we compare a cavity with slightly increased thickness (~0.1%) with the unmodified structure. The parameters considered in this analysis are as follows: 1) reduction in organic dye layer absorption, to replicate the effect of ground-state bleaching; 2) thickness of the organic layer (keeping the total film absorption constant), to capture thermal expansion; and background refractive index changes in 3) the organic polymer matrix and 4) the mirror material (metal or dielectric), in accord with previous transient studies on photonic structures.[78,79] Within this scheme we only calculate steady-state optical properties; our assumption is that any spectral dynamics arise from the mechanism of the underlying effect. For instance, within the two-temperature model of Liu *et al.*, the <10 ps dynamics in microcavity transient reflectivity could be predominantly described by carrier thermalization.[63] Similar dynamics are reported for hot carrier cooling in other photonic structures.[78,79] Ground-state bleaching should roughly follow the dynamics of the bare film, resulting in persistent signatures on the order of nanoseconds. These effects span the sub-ps to ns timescale and could easily be reconciled with our measured dynamics. Thus, we do not explicitly treat the time dependence of these effects here, and our principal concern is the spectral shapes that arise. In the discussion below we focus on transmission data for ease of comparison with our experimental results, but the same basic behaviour is captured in the modelled reflectance (Supplementary Information).

The full angle dependence of our non-specific pump-induced effects is presented in Figure 5 for perturbation magnitudes chosen to reproduce typical experimentally observed ΔT/T signal strengths. All reveal prominent derivative-like features centred on the upper and lower polariton resonances. These are qualitatively similar to the features often reported in organic microcavity transient absorption experiments and assigned to polaritons.[36,50–52] Interestingly, we additionally calculate clear spectral signatures at the positions of the DBR sidebands, even when the effect we consider has no relation to the mirrors such as bleaching of the electronic ground state (Figure 5a).

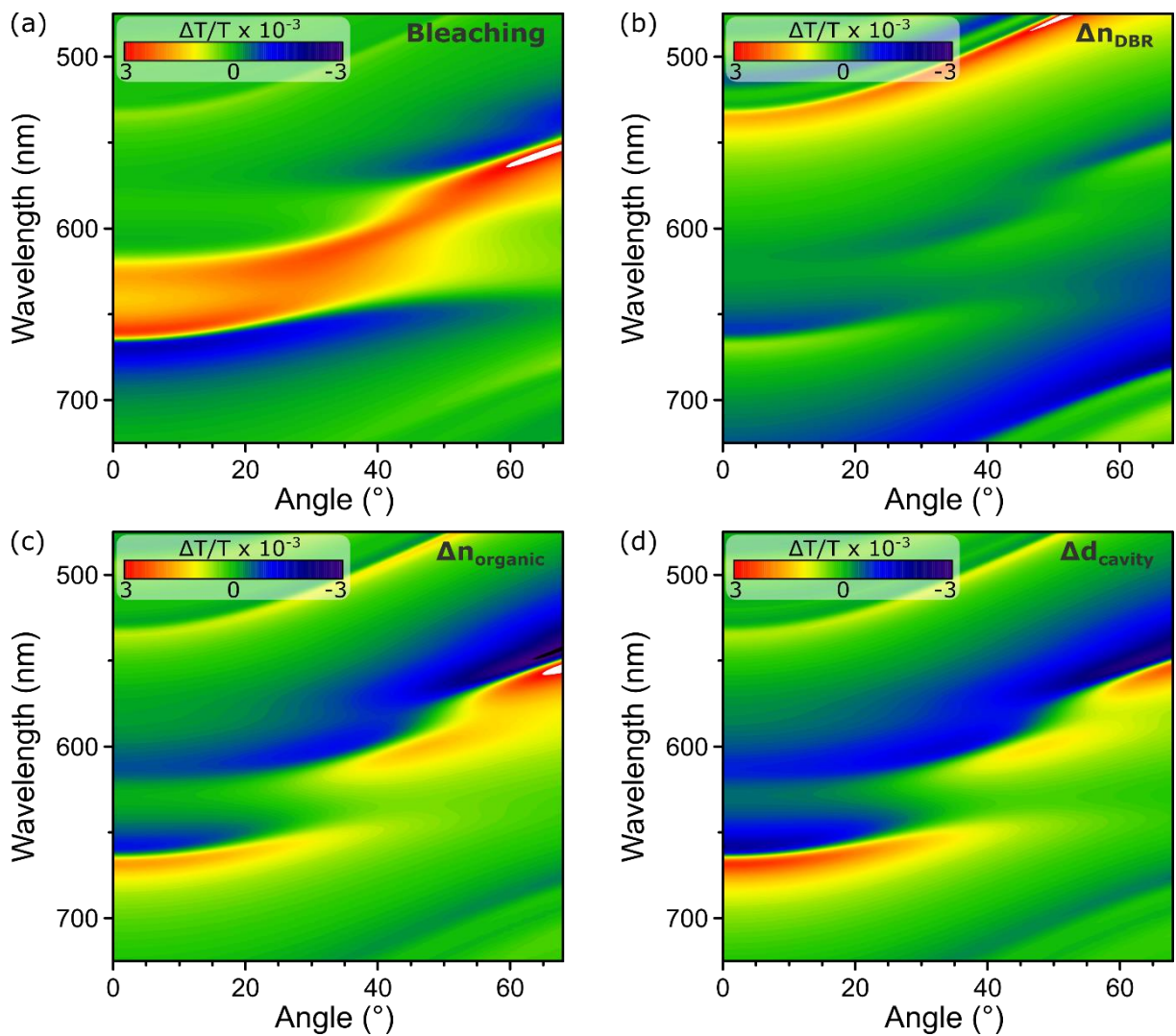

**Figure 5. Transmission changes due to non-specific photoinduced effects in microcavities.** Transient transmission maps calculated for **(a)** ground-state bleaching of 0.1% of population; **(b)** a uniform increase in DBR refractive by 0.001%; **(c)** a uniform Increase in refractive index of the organic layer by 0.01%; and **(d)** an increase in the thickness of the organic layer by 0.005%.

The effect of bleaching has been previously considered, particularly in the field of vibro-polaritons, and described in terms of polariton contraction.[58,59] The derivative-like lineshape we recover results in a blue-shift in transmittance at low angle and a red-shift in transmittance at high angle, reflecting a narrowing of the gap between upper and lower polariton branches. This is an expected result from the reduction in the number of absorbing molecules in the ground state, which is related to the Rabi splitting as $\Omega \propto \sqrt{N}$. This can yield pronounced transient effects even at relatively small excitation density. The model in Figure 5a assumes 0.1% of the molecules are photoexcited, a typical degree of excitation in transient absorption experiments, and the resulting signal changes are likewise within the range measured by ourselves and others.[36,50–52,62,63] In previous studies, the contribution of such polariton contraction has been ruled out through intensity-dependence measurements, where it is assumed that the $\Omega \propto \sqrt{N}$ relation should yield progressively smaller Rabi splitting and thus increasingly shifted polariton resonances as the laser power is increased.[52,62] Our calculations reveal that the assumption behind this approach in incorrect, and intensity dependence cannot be used for this determination (Figure 6). The spectral shape arising from the bleaching effect remains almost identically the same, and the magnitude scales almost precisely linearly with the degree of excitation, up to >1% bleaching. This behavior is a straightforward consequence of the fact that the bulk film absorption is only slightly altered in typical transient absorption experiments, and that slight alteration is further mitigated through the $\sqrt{N}$ dependence and the linewidths attainable in our experiments. Indeed, only for extreme excitation fractions of ~10% or greater is the deviation from linearity large enough to potentially be detected in transient absorption experiments. In short, the reported absence of an intensity dependence to the spectral shape[52,62] does not rule out contributions from bleaching effects.

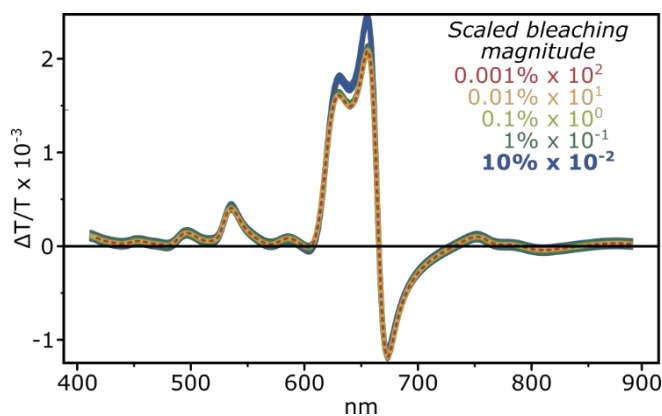

**Figure 6. Linearity of polariton contraction.** Simulated transient transmission through a BODIPY-R/DBR cavity for the degree of bleaching indicated. Each spectrum is scaled by excitation fraction. Only at very high powers, beyond the regime of reliable transient absorption measurements, do deviations in shape become detectable.

The remaining three effects, when modelled as an increase in *n* or *d*, produce a red-shift across the entire angular range, principally due to a shift in the energy of the uncoupled cavity mode (Figure 5b-d). The changes in refractive index within the DBR or organic layer are qualitatively similar, though they can be distinguished at higher angles where the curvature of the cavity mode is prominently affected by $n_{organic}$. The effect of $n_{DBR}$ is especially pronounced outside of the cavity stop-band through modulations of the mirror side-bands, providing multiple points to uniquely distinguish these effects. We note that markedly smaller spectral effects are obtained from an equivalent change in the Ag refractive index in BODIPY/Ag cavities (Supplementary Information), which we attribute to the substantially thinner Ag layer in these structures.

In Figure 7 we plot spectral cuts of our calculated transient transmission maps at low angle for DBR cavities. The magnitude of simulated perturbations is small and easily achieved within a transient experiment (e.g. 0.1% excitation fraction, or Å-scale thickness changes in a 200-nm film), yet the signal magnitudes are easily comparable with experimental results. Thus, these features must be understood and explicitly considered in the interpretation of transient absorption spectroscopy on strongly coupled organic microcavities. Moreover, it is likely that all of these will be present at the same time. We have explored the linearity of all effects, both in terms of individual scaling as for polariton contraction above and for combinations of different effects (see Supplementary Information for details). We find that within the ranges relevant to this study (where the peak ΔT/T<0.01), all four effects scale linearly with perturbation magnitude and can be linearly combined with negligible errors. Thus, we can treat the spectra in Figure 7 as basis spectra to approximate our transient transmission data.

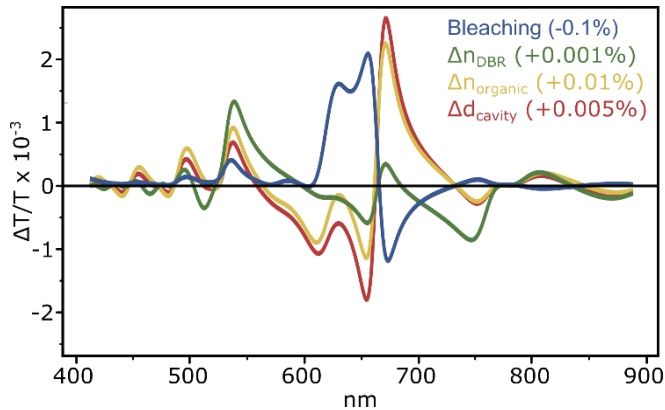

**Figure 8. Transient transmission spectra of non-specific photoinduced effects.** Spectral cuts from the maps in Figure 5, taken at low angle. The small perturbations indicated are sufficient to generate experimental-scale ΔT/T signatures. Due to their linearity, these spectra can be used as basis spectra to model full transient transmission data.

The best fit to the data at 1 ps, shown in Figure 8, includes 0.03% bleaching, 0.0024% decrease in refractive index of the DBRs, 0.008% increase in the refractive index of the organic layer, and a 0.0085% expansion in cavity thickness. Because of the close spectral similarity of organic layer refractive index and thickness effects (Figure 7) the solution at a single angle is not uniquely determined. Additional measurements at higher angles would enable disentanglement of these contributions. Nevertheless, this simple qualitative model reproduces many of the key features highlighted in Figure 3, particularly those that appear in the absence of electronically excited states (Figure 3e). Even the modulations in the NIR appear at the correct spectral position. The fact that we can capture so much of the rich structure of our transient experiment without any explicit consideration of intracavity or polaritonic excited states suggests that the transient transmission data is dominated by non-specific optical effects. We note that the character of the fit is extremely sensitive to the film thickness. Increasing the baseline active layer thickness by 10 nm results in excellent agreement with the central absorptive peaks ~700 nm at the expense of the absorptive band at 650 nm (Supplementary Information). Given such thickness variations are common in organic thin films, including similar BODIPY cavities,[61] quantitative accounting for these effects likely demands more uniform, evaporated cavity layers. Moreover, our modelled spectrum in Figure 8 significantly diverges from the experimental result at long wavelengths. This difference should be chiefly related to the contribution of the transient absorption of intracavity BODIPY-R excitons, i.e. states within the exciton reservoir, giving rise to an extended PIA in the near-infrared and complex features in the GSB region. Similarly, the slight positive peak in the experimental data at 710 nm matches the BODIPY-R stimulated emission (Figure 3f). Though not directly incorporated into our scheme, these signatures can be readily identified.

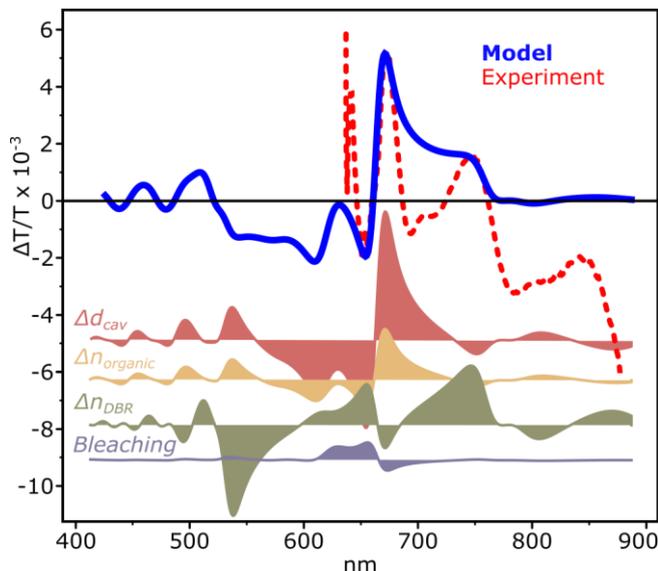

**Figure 7. Phenomenological simulated transient transmission spectrum.** Best-fit model of experimental data, composed from the following sample changes: 0.03% bleaching, 0.0085% increased thickness, 0.0024% decrease in the DBR refractive index, 0.008% increase in the organic refractive index. Experimental data is reproduced from Figure 3b, 1 ps delay. Individual components are offset to highlight their relative magnitudes.

These transfer-matrix calculations and the persistence of the polariton-like spectral signatures even in the absence of excited electronic states suggest that the dominant transient transmission features in Figures 2,3 do not reflect long-lived exciton-polaritons. This contradicts earlier findings in other strong-coupled systems.[51,52,75,76] However, any given photoexcited polariton in our and others' microcavities should have a lifetime chiefly limited by the cavity photonic lifetime, through $\tau_{pol}^{-1} = A\tau_p^{-1} + B\tau_x^{-1}$, where $\tau_p$ and $\tau_x$ are the lifetimes of the unmixed photonic and excitonic states and $A$ and $B$ their respective fractions in the wavefunction. The Q-factors of our cavities thus give an upper limit to the polariton lifetime of 5 fs (Ag) or 110 fs (DBR), well within the resolution of our measurements. This limitation is common to all reported organic exciton-polariton transient absorption studies[36,51–54] except for the first,[50] which is also the only study to directly distinguish the features of reservoir excitons from polaritons. While long-lived intracavity excitons can serve as an exciton reservoir and populate the LP over ps, ns and even µs timescales, rapid intrinsic losses from the LP mean k$_{scattering}$<<k$_{LP\_decay}$ and this process is incapable of producing a detectable LP population signal. In the absence of a clear model to explain long-lived polaritonic states, we find that the most plausible explanation of the spectra we detect is a combination of intracavity excitonic states,[62] modification of the Rabi splitting through depopulation of the ground state and non-specific thermal effects including transient changes to the bulk refractive indices and thermal expansion.

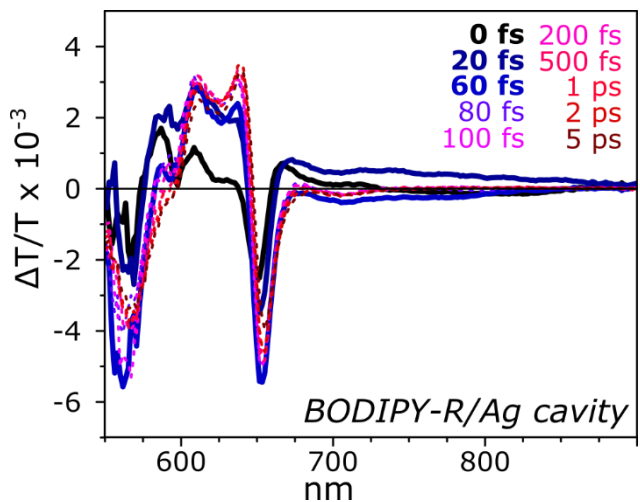

**Figure 9. Ultrafast polariton signatures.** Transient transmission of a BODIPY-R cavity with Ag mirrors, excited with a broadband, 15 fs pump pulse. By 80 fs (dashed) the spectrum has assumed the same shape detected on longer timescales.

The implication of this finding is that to directly observe the spectral signatures and dynamics of the UP and LP states, we must use excitation pulses with duration comparable to or shorter than the state lifetime. Accordingly, we have performed preliminary broadband transient transmission measurements on BODIPY-R/Ag microcavities using <15-fs excitation pulses, shown in Figure 9. By 80 fs time delay, the spectrum has assumed the same shape reported in slower measurements and discussed above. Only in the first spectral slices (bold) do we detect any features distinct from the long-time spectrum, which we infer from the above analysis to consist chiefly of non-specific optical effects and reservoir excitons. These features are distinct from the coherent response of the instrument and indicate unique ultrafast dynamics, but their interpretation is beyond the scope of the present study. Nonetheless, the results suggest ultrafast spectroscopy combined with higher-Q microcavities holds the promise to enable direct identification of the unique spectral features of exciton-polaritons and elucidation of their dynamics, a critical step towards unlocking their potential to modify photochemistry.

**Conclusions**

We have explored the time-resolved spectroscopy of strongly coupled microcavities containing the organic dye BODIPY-R. Our transient transmission data revealed many of the same types of features—chiefly derivative-like lineshapes centred at the polariton resonances—described in previous organic polariton systems.[33,36,50–52,54,56] However, the lifetimes of these features were strikingly longer than what would be reasonably expected based on the partially photonic character of the polariton states. We consider this behaviour to be a common, and commonly overlooked, red flag that the spectral features cannot be simply assigned to photoexcited exciton-polaritons. Indeed, these features were accompanied by similar modulations at the stop-band edge and DBR sideband, which are highly unlikely to directly relate to exciton-polaritons. Moreover, we find that many of the same underlying spectral features are present whether excitation is above, at, or below the polariton resonance. On the basis of transfer matrix modelling, we propose that these spectral features and their unexpectedly long lifetimes can be explained through non-specific photoinduced effects such as polariton contraction. We find that numerous small changes which could reasonably be expected to occur in any transient transmission experiment readily yield spectral signatures of the same magnitude as those reported here and elsewhere, [36,50–54] and the lifetimes of these effects can span the ps-ns regimes. As highlighted by Liu *et al*.[62,63] and in parallel studies of vibropolaritons,[57–60] the Fabry-Perot structures used in this and similar works are complex optical systems despite their relatively simple structure. Accordingly, their optical spectra are highly sensitive to changes in the optical properties of any one of the components. These effects can occur regardless of the nature of the material composing the cavity mirrors or active layer, and their linearity with degree of excitation means they can be challenging to separate from real excited-state effects. We thus conclude that similar effects are likely present in all previous transient absorption and related studies of organic exciton-polaritons in Fabry-Perot cavities,[36,50–55] and that these effects may explain the altered photophysical dynamics and surprisingly long lifetimes that have been reported.

While the many overlapping contributions from these non-specific effects pose a substantial challenge for rigorous exciton-polariton spectroscopy, our results suggest a fruitful path forward. The unique angle dependence of each contribution in Figure 5 is a potent tool to disentangle their signatures. We would expect any real signatures of polaritons to exhibit still different angle dependence, whether they continuously relax down the lower polariton branch or remain stuck at a bottleneck. Just as the angle dependence of photoluminescence provides important insight into the polariton population distribution and dynamics,[41–45] we expect that angle-dependent transient absorption will prove necessary to unambiguously separate the multiple species contributing to the transient absorption dynamics. This is rarely probed and is not systematically explored at present, and there is as yet no equivalent of the widespread Fourier-plane imaging used to record polariton emission. The other crucial tool already available is temporal resolution. To unambiguously assign the dynamics of organic exciton-polaritons, it is essential to employ a temporal resolution shorter than the expected polariton lifetime. This requires the use of compressed, ultrafast pulses and/or high-Q dielectric cavities, as combined in the original study of Virgili *et al.* but not applied since.[50] Through this combination and careful accounting for non-specific optical effects, we anticipate it will be possible to uncover the intrinsic dynamics and rich photochemistry of organic molecules in the strong-coupling regime.

**ACKNOWLEDGEMENTS**

This work was supported by by the Engineering and Physical Sciences Research Council, U.K. (Grant Number EP/M025330/1, 'Hybrid Polaritonics').